\documentstyle[sprocl]{article}
\input{epsf}

\def\be{\begin{equation}}
\def\ee{\end{equation}}
\def\bea{\begin{eqnarray}}
\def\eea{\end{eqnarray}}
\begin{document}

\def\question#1{{{\marginpar{\small \sc #1}}}}
\def\mpl{{{m_{Pl}}}}
\def\oph{{{\Omega_{\widetilde{\gamma}}h^2}}}
\def\be{\begin{equation}}
\def\ee{\end{equation}}
\def\ba{\begin{eqnarray}}
\def\ea{\end{eqnarray}}
\def\la{\mathrel{\mathpalette\fun <}}
\def\ga{\mathrel{\mathpalette\fun >}}
\def\fun#1#2{\lower3.6pt\vbox{\baselineskip0pt\lineskip.9pt
        \ialign{$\mathsurround=0pt#1\hfill##\hfil$\crcr#2\crcr\sim\crcr}}}
\def\pho{{{\widetilde{\gamma}}}}
\def\r0{{{R^0}}}
\def\glu{{{\widetilde{g}}}}
\def\sec{{{\mbox{sec}}}}
\def\GeV{{{\mbox{GeV}}}}
\def\MeV{{{\mbox{MeV}}}}
\def\SUSY{{{{\sc susy}}}}
\def\LSP{{{{\sc lsp}}}}
\def\LEP{{{{\sc lep}}}}
\def\LROCS{{{{\sc lrocs}}}}
\def\WIMPS{{{{\sc wimps}}}}
\def\cm{{{\mbox{cm}}}}
\def\photino{{{\mbox{photino}}}}
\def\gluino{{{\mbox{gluino}}}}
\def\mb{{{\mbox{mb}}}}
\def\avg#1{{{{\langle #1 \rangle }}}}
\def\taun{{{\tau_{9}}}}
\def\mpl{{{m_{Pl}}}}
\def\re#1{{[\ref{#1}]}}
\def\eqr#1{{Eq.\ (\ref{#1})}}
\def\mst{{{M_{\widetilde{S}}}}}
\newcommand{\gl}{\tilde{g}}
\newcommand{\sneu}{\tilde{\nu}}
\newcommand{\sq}{\tilde{q}}
\newcommand{\se}{\tilde{e}}
\newcommand{\ch}{\chi^{\pm}}
\newcommand{\neut}{\chi^{0}}
\newcommand{\gsi}{\,\raisebox{-0.13cm}{$\stackrel{\textstyle>}
{\textstyle\sim}$}\,}
\newcommand{\lsi}{\,\raisebox{-0.13cm}{$\stackrel{\textstyle<}
{\textstyle\sim}$}\,}

\title{Dark Matter and Cosmic Rays from Light Gauginos}
\author{G. R. Farrar}

\address{Department of Physics and Astronomy \\ Rutgers University,
Piscataway, NJ 08855, USA}  

\maketitle\abstracts{
An attractive class of SUSY-breaking mechanisms predicts a photino mass
of order 1 GeV.  Relic photinos can naturally account for the observed
dark matter.  Detection of these light photinos is discussed and
contrasted with conventional WIMPs.  In this scenario the gluino mass
is about 100 MeV.  The lightest gluino-containing baryon could account
for the recently observed ultra-high energy cosmic rays, which violate
the GZK bound. 
}
Some supersymmetry (SUSY) breaking scenarios produce negligible
tree-level gaugino masses and negligible scalar trilinear couplings
($M_1=M_2=M_3=A=0$) and conserve $R$-parity.  Such SUSY breaking
has several attractive theoretical consequences such as the
absence of the ``SUSY CP problem"\cite{f99101}.  Although
massless at tree level, gauginos get calculable masses through
radiative corrections from electroweak (gaugino/higgsino-Higgs/gauge
boson) and top-stop loops. Evaluating these within the constrained
parameter space leads to a gluino mass range $m_{\tilde{g}}\sim
\frac{1}{10} - \frac{1}{2}$ GeV\cite{f99101}, while analysis of the
$\eta'$ mass narrow this to $m(\glu) \approx 120$ MeV\cite{f:108}. The
photino mass range depends on more unknowns than the gluino mass, such
as the higgs and higgsino sectors, but can be estimated to be
$m_{\tilde{\gamma}} \sim \frac{1}{10} - 1 \frac{1}{2}$
GeV\cite{f99101}.   

The gluino binds with quarks, antiquarks and/or gluons to make
color-singlet hadrons (generically called $R$-hadrons\cite{f:24}). 
The lightest of these is expected to be the gluino-gluon bound state,
designated $\r0$.  It is predicted to have a mass in the range
$1.3-2$ GeV, approximately degenerate with its superpartners the
lightest glueball ($0^{++}$) and ``gluinoball'' ($0^{-+},~\glu
\glu$)\cite{f:95,f:104}.  An encouraging development for this scearnio is
the experimental evidence for an ``extra'' isosinglet pseudoscalar
meson, $\eta(1410)$, which is difficult to accomodate in standard QCD
but which matches the properties of the pseudoscalar $\glu
\glu$\cite{f:109,f:104}.  

Due to the non-negligible mass of the photino compared to the $\r0$,
the $\r0$ is long lived.  Its lifetime is estimated to be in the range 
$10^{-10} - 10^{-5}$ sec\cite{f99101,f:104}.  Prompt
photinos\cite{f:24} are not a useful signature for the light gluinos
and the energy they carry\cite{f:51}. Thus gluino masses less than
about $ \frac{1}{2}$ GeV are largely
unconstrained\cite{f:95}\footnote{The recent ALPEH claim to exclude 
light gluinos[\ref{aleph:lg}] assigns a $1 \sigma$ theoretical
systematic error based on varying the renormalization scale over a
small range. Taking a more generally accepted range of scale variation
and accounting for the large sensitivity to hadronization model, the
ALEPH systematic uncertainty is comparable to that of other
experiments and does not exclude light gluinos[\ref{fLaT}].}.
Proposals for direct searches for hadrons containing gluinos, via
their decays in $K^0$ beams and otherwise, are given in Refs.
\cite{f:95,f:104}.  It is noteworthy that this ``light gaugino
scenario'' can naturally account for the mass peak observed in $e^+
e^- \rightarrow 4$ jets by ALEPH (but not other LEP experiments) at
$E_{cm}$ of 130-136 and 161 GeV\cite{aleph:4j,f:105,f:112}.  

In the light gaugino scenario, photinos remain in thermal equilibrium
much longer than in conventional SUSY, due to pion catalysis of their
conversion to $R^0$'s: $\pho \pi \leftrightarrow \r0 \pi$.  The
$\r0$'s stay in thermal equilibrium still longer, because their
self-annihilation to pions has a strong interaction cross section.
The relic abundance of photinos depends sensitively on the ratio of
the $R^0$ and $\pho$ masses, $M$ and $m$, respectively\cite{f:100}.
This is because the Boltzman probability of finding a pion with
sufficient energy to produce an $\r0$ from a $\pho$ decreases
exponentially as the $\r0$ mass increases.  This was studied in sudden
approximation in ref. \cite{f:100} using the most relevant
reactions.  The analysis has been refined (see ref. \cite{f:113}) by
integrating the coupled system of Boltzman equations for the reactions
$\pho \pi \leftrightarrow \r0 \pi$, $\r0 \leftrightarrow \pi^+ \pi^-
\pho$, $\r0 \pho \leftrightarrow \pi^+ \pi^- $, and $\r0 \r0$ total
annihilation.  Defining $r \equiv \frac{M}{m}$, ref. \cite{f:113}
finds that for relic photinos to give $\Omega h^2 \sim 0.25$ requires
$1.2 \la r \la 1.8$.  This range of $r$ is consistent with the mass
estimates quoted above for the $\r0$ and $\pho$, which encourages to
take the possibility of light photinos seriously.

The detectability of relic dark matter is different for light $\pho$'s
than in the conventional heavy WIMP scenario for two reasons.
The usual relation between the relic density and the
WIMP-matter scattering cross section only applies when the relic
density is determined by the WIMP self-annihilation cross section.  In
order to have the correct relic abundance, the rate of photino-removal
from the thermal plasma must be greater than the expansion rate of the
Universe until a temperature of order $m/22$.  In the light photino
scenario, the photino relic density is determined by the $\r0 - \pho$
interconversion cross section.  When the dominant process keeping
photinos in thermal equilibrium is interconversion, whose rate $\sim
n_\pi <\sigma_{R \pi \leftrightarrow \pho \pi} v>$ the required 
interconversion cross section $\sigma_{R \pi \leftrightarrow \pho
\pi}$ is smaller than the required $\sigma_{\pho \pho \leftrightarrow f
\bar{f}}$ when the rate governing photino equilibrium is $n_\pho
<\sigma_{\pho \pho \leftrightarrow f \bar{f}} v>$, because $n_\pi >>
n_\pho$\cite{f:100}.  The photino-matter scattering cross section is
therefore correspondingly smaller as well.  Goodman and Witten in Ref.
\cite{goodwitt} discuss $\pho$ detection through $\pho$-nucleon
elastic scattering. Using Eq.\ (3) of Ref. \cite{goodwitt} and the
parameters for light photinos, one finds event rates between $10^{-3}$
and $10$ events/(kg day)\cite{f:113}.    

Even if the event rate were larger, observation of relic light
photinos would be difficult with existing detectors because the
sensitivity of a generic detector is poor for $\la 1$ GeV mass
relevant in this case, because WIMP detectors have generally been 
optimized to maximize the recoil energy for a WIMP mass of order $10$
to $100$ GeV.   

\begin{figure}
\hspace*{25pt} \epsfxsize=400pt \epsfbox{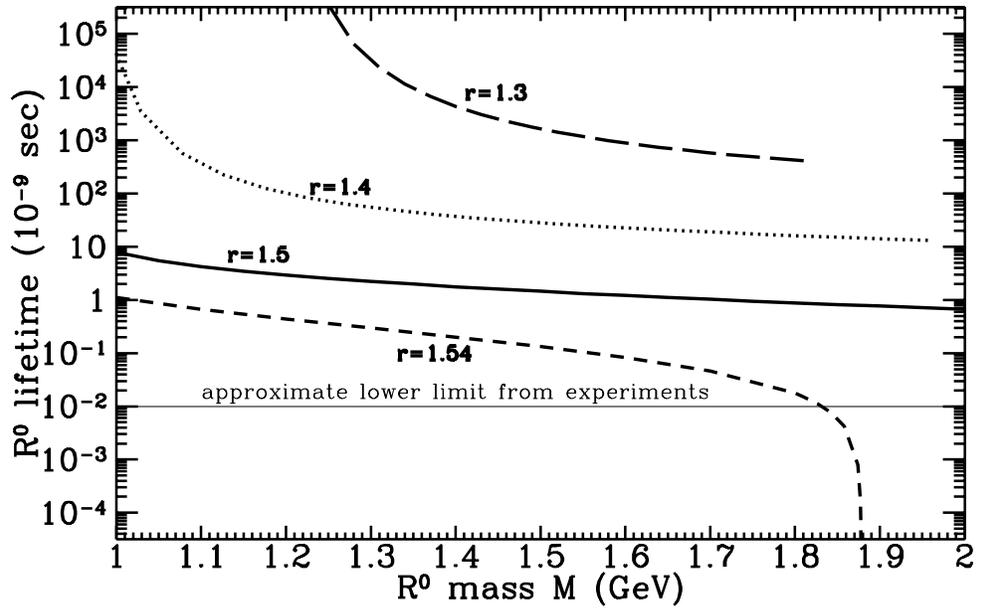}\\
\caption{$R^0$ lifetime (strictly speaking, upper limit thereto) as a
function of the $R^0$ mass, for several values of $r \equiv
M_\r0/m_\pho$, when $\Omega_\pho h^2 = 0.25$. }
\label{fig:tau}
\end{figure}

The amplitudes for the reactions responsible for $R^0$ decay
($\r0 \leftrightarrow \pi^+ \pi^- \pho$) and the photino 
relic density ($\pho \pi \leftrightarrow \r0 \pi$) are related by
crossing symmetry.  If the momentum dependence of the amplitude is 
mild, the $\r0$ lifetime and the photino relic abundance depend
on a single common parameter in addition to the $\r0$ and $\pho$
masses $M$ and $m$\cite{f:113}.  In that case, demanding the correct
dark matter density determines the $\r0$ lifetime given the 
$\r0$ and $\pho$ masses.  The resulting lifetimes are shown in Fig.
\ref{fig:tau} from \cite{f:113}.  In actuality, the interconversion
reaction $\pho \pi \leftrightarrow \r0 \pi$ is expected to have a
resonance, so momentum-independence of the amplitudes is not a good
assumption for all of parameter space.  However this merely lengthens
the $\r0$ lifetime in comparison with the crossing relation Fig.
\ref{fig:tau}.  The required lifetime range is consistent with both
the experimental limits\cite{f:95} and with the predicted range of
lifetimes, so relic light photinos pass an important hurdle.
Experiments currently underway should be sensitive to much of the
lifetime range of interest\cite{f:104}. 

In the light gaugino scenario is correct, the lightest $R$-baryon,
$S^0 \equiv uds\glu$, may be responsible for the very highest energy 
cosmic rays reaching Earth.  Recall that the observation of several
events with energies $\ga 2~ 10^{20}$ eV\cite{akeno_flyseye} presents
a severe puzzle for astrophysics\footnote{For a recent survey and
references see \cite{uhecr}.}.  Protons with such high energies have a
large scattering cross section on the 2.7 K microwave background
photons, because $E_{cm}$ is sufficient to excite the $\Delta(1230)$
resonance\cite{GZK}.  Consequently the scattering length
of such high energy protons is of order 30 mpc or 
less.  The upper bound on the energy of cosmic rays which could have
originated in the local cluster, $\sim 10^{20}$ eV, is called the
Greisen-Zatsepin-Kuzmin (GZK) bound.   

\begin{figure}
\hspace*{25pt} \epsfxsize=400pt \epsfbox{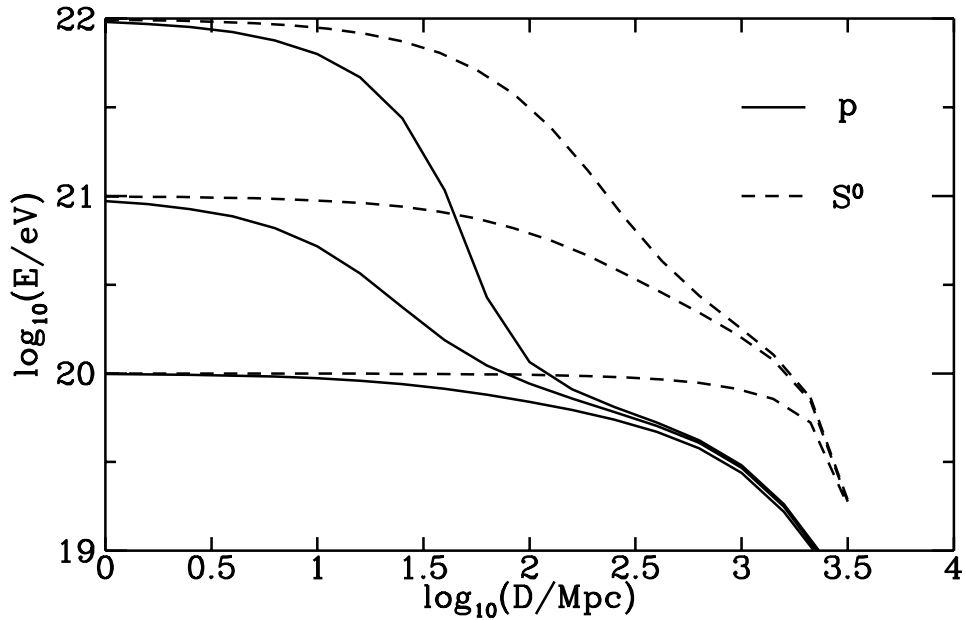}\\
\caption{Energy of a high energy proton and $S^0$ as a function of
distance from source, for various initial energies.}
\label{gzk}
\end{figure}

Two of the highest energy cosmic ray events come from the same
direction in the sky\cite{uhecr}.  The nearest plausible source in
that direction is the Seyfert galaxy MCG 8-11-11 (aka UGC 03374), but
it is 62-124 Mpc away\cite{ElbSom}.  Fig. \ref{gzk} (taken from 
\cite{f:114}) shows the spectrum of high energy protons as a function
of their initial distance, for several different values of the energy.
Compton scattering and photoproduction, as well as redshift effects,
have been included.  Evidently, it is unlikely that the highest energy
cosmic ray events can be due to protons from MCG 8-11-11, and even
more unlikely that two high energy protons could penetrate such
distances.  

However the ground-state $R$-baryon, the flavor singlet scalar $uds
\tilde{g}$ bound state denoted $S^0$, could explain these
ultra-high-energy events\cite{f:104}.  On account of the very 
strong hyperfine attraction among the quarks in the flavor-singlet
channel\cite{f:52}, the $S^0$ mass is about $210 \pm 20$ 
MeV lower than that of the lowest $R$-nucleons.  As long as $m(S^0)$ is
less than $m(p) + m(R^0)$, the $S^0$ must decay to a photino rather
than $R^0$.  It would have an extremely long lifetime since its decay
requires a flavor-changing-neutral-weak transition.  The $S^0$ could
even be stable, if $m(S^0) - m(p) - m(e^-) < m_{\tilde{\gamma}}$ and
baryon number is a good quantum number\cite{f:104}.  

The GZK bound for the $S^0$ is several times higher than for
protons. Three effects contribute to this:  (a) The $S^0$
is neutral, so its interactions with photons cancel at leading order
and are only present due to inhomogeneities in its quark substructure.
(b) The $S^0$ is heavier than the proton.  (c) The mass splitting
between the $S^0$ and the lowest lying resonances which can be reached
in a $\gamma S^0$ collision (mass $\equiv M^*$) is larger than the 
proton-$\Delta(1230)$ splitting.  

The threshold energy for exciting the resonances in $\gamma S^0$
collisions is larger than in $\gamma p$ collisions by the  
factor\cite{f:104} $\frac{m_{S^0}}{m_p} \frac{( M^* -
M_{S^0})}{(1230-940) {\rm MeV}}$.  We can estimate $M_{S^0}$ and $M^*
- M_{S^0}$ as follows.  Taking $m(R^0) = 1.7$ GeV,
$m_{\tilde{\gamma}}$ must lie in the range $0.9 \sim 1.4$ GeV to
account for the relic dark matter.  If $m_{S^0} \approx m_p + 
m_{\tilde{\gamma}}$ we have $m(S^0) \sim 1.9 - 2.3$ GeV. 
Since the photon couples as a flavor octet, the resonances excited in
$S^0 \gamma$ collisions are flavor octets.  Since the $S^0$ has
spin-0, only a spin-1 $R_{\Lambda}$ or $R_{\Sigma}$ can be produced
without an angular momentum barrier.  There are two $R$-baryon flavor
octets with $J=1$, one with total quark spin 3/2 and the other with
total quark spin 1/2, like the $S^0$.  Neglecting the mixing between
these states which is small, their masses are about 385-460 and
815-890 MeV heavier than the $S^0$, respectively\cite{f:52}.  Thus one
qualitatively expects that the GZK bound is a factor of 2.7 - 7.5
higher for $S^0$'s than for $p$'s, depending on which $R$-hyperons are
strongly coupled to the $\gamma S^0$ system.  

A more detailed calculation of $S^0$ scattering on microwave 
photons will soon be published\cite{f:114}.  The results for a typical
choice of parameters are shown in Fig. \ref{gzk}, confirming the
crude treatment of ref. \cite{f:104}.  

If $S^0$'s are stable they naturally increase the GZK bound enough to
be compatible with the highest energy cosmic rays reported up to now
(see \cite{akeno_flyseye} and references therein).  However it is
enough that the $S^0$ lifetime be longer than $\sim 10^5$ sec.  This
is the proper time required for a few $10^{20}$ GeV particle of mass
$\sim 2$ GeV to travel $\sim 100$ Mpc.

With the much larger sample of ultra-high-energy cosmic rays expected
from the Auger Project, the prediction of a GZK cutoff shifted to
higher energy can hopefully be tested.  Furthermore, the $S^0$'s are
not deflected by magnetic fields so they should accurately point
to their sources.  Indeed, being neutral spin-0 particles, even their
magnetic dipole moment vanishes.

To summarize:\\
$\bullet$ Light photinos can account for the relic dark matter, if the
$R^0$ mass is between 1.2 and 1.8 times the photino mass.  This is
consistent with the predicted mass ranges.  \\ 
$\bullet$ If the dark matter is due to relic photinos, one can expect
$10^{-3} - 10$ interactions per kg per day.  However since the photino
mass is of order 1 GeV in this scenario, they will not deposit 
significant energy in detectors based on heavy nuclei. \\
$\bullet$  The cosmic ray events whose energy is above the GZK bound
may be due to the lightest gluino-containing baryon, a $uds\gl$
bound state called the $S^0$.  A cutoff in the spectrum at a somewhat
higher energy is predicted, as is sharp pointing to the sources.

\section*{Acknowledgments}
Researh supported in part by NSF-PHY-94-23002.  Much of the research
reported here was done in collaboration with D. J. Chung and E. W.
Kolb.  I thank them for enjoyable and productive collaboration, 
and permission to reproduce figures from our as-yet-unpublished work.

\section*{References}

\end{document}